\documentclass[twocolumn,showpacs,showkeys,preprintnumbers,groupedaddress,aps,amsmath,amssymb,prb,a4paper]{revtex4}
\usepackage{graphicx}% Include figure files
\usepackage{dcolumn}% Align table columns on decimal point
\usepackage{bm}% bold math
\usepackage{natbib}
\newcommand{\ycs}{YbCo$_{2}$Si$_{2}$}
\newcommand{\yrs}{YbRh$_{2}$Si$_{2}$}
\newcommand{\yrcs}{Yb(Rh$_{1-x}$Co$_{x})_{2}$Si$_{2}$}
\newcommand{\et}{~\textit{et al.~}}
\begin{document}
\preprint{APS/123-QED}
\title{Magnetic phase diagram of YbCo$_2$Si$_2$ derived from magnetization measurements}
\author{L. Pedrero, C. Klingner, C. Krellner, M. Brando, C. Geibel and F. Steglich}
\affiliation{Max Planck Institute for Chemical Physics of Solids, N\"othnitzer Str. 40, Dresden 01178, Germany}
\email{manuel.brando@cpfs.mpg.de}
\date{\today} 
\begin{abstract}
We report on high-resolution dc-magnetization ($M$) measurements on a high-quality single crystal of \ycs. $M$ was measured down to 0.05\,K and in fields up to 3\,T, with the magnetic field oriented along the crystallographic directions
$[100]$, $[110]$ and $[001]$ of the tetragonal structure.\\
In a small field $\mu_{0}H = 0.1$\,T two antiferromagnetic (AFM) phase transitions have been detected at $T_{N} = 1.75$\,K and $T_{L} = 0.9$\,K, in the form of a sharp cusp and a sudden drop in $\chi = M/H$. These signatures confirm that the phase
transitions are $2^{nd}$ order at $T_{N}$ and $1^{st}$ order at $T_{L}$. The AFM order is completely suppressed by fields as high as $\mu_{0}H_{N}(0)=1.9$, 1.88 and 2.35\,T for $H // [100]$, $[110]$ and $[001]$, respectively. At these fields, $M$
reaches its saturation values $M_{s}(H // [100]) \approx 1.4 \mu_{B}$ and $M_{s}(H // [001]) \approx 0.68 \mu_{B}$ which match quite well those calculated for the $\Gamma_{7}$ ground state proposed in Ref.~\onlinecite{Klingner2010a} and confirm the
trivalent state of the Yb ions in \ycs.\\
We have derived the $H - T$ phase diagrams along the three crystallographic directions from isothermal and isofield measurements. Four AFM regions can be identified for $H$ parallel to $[100]$ and $[110]$ which are separated by $1^{st}$ and $2^{nd}$
order phase-transition lines, showing anisotropy in the basal plane. For $H$ parallel to $[001]$ only two AFM phases have been observed. The phase boundary $T_{N}(H)$, which separated the AFM phase from the paramagnetic one, can be well described
by the empirical relation $[H_{N}(T)/H_{N}(0)]^{n}+[T/T_{N}(0)]^{n} = 1$ with $n$ close to 2. 
\end{abstract}
%\keywords{\ycs,\yrs}%Use showkeys class option if keyword
\maketitle
\section{Introduction}
In Yb-based intermetallic compounds magnetic ordering is relatively rare, since usually the Yb atom has a valence close to +2, which implies a non-magnetic ground state. In few rare cases the Yb valence is close to three. Then the Yb$^{3+}$ local magnetic
moments can both interact with the conduction electrons (Kondo effect) and with each other through the Ruderman-Kittel-Kasuya-Yoshida (RKKY) exchange interaction.~\cite{Stewart1984} One possibility is that a strong Kondo effect can progressively screen the Yb magnetic moments
leaving a non-magnetic Fermi-liquid ground state. If the Kondo effect is not strong enough, however, magnetic order can be established by the RKKY interaction, and in Yb-based intermetallics it is quite often antiferromagnetic (AFM). 
A special case is \yrs, which, with a Kondo temperature $T_{K}\approx 25$\,K and an antiferromagnetically ordered state below $T_{N} = 0.07$\,K, is close to the critical point where the AFM ground state gets suppressed by the increasing
Kondo interaction.~\cite{Trovarelli2000b} A further sharp phase transition has been observed in magnetization measurements at a temperature $T_{L} = 0.002$\,K;~\cite{Schuberth2009}
its origin is still unclear, but the comparison with the homologue \ycs~\cite{Pedrero2010a} and the evolution of $T_{L}$ observed in the series \yrcs\, suggest a second AFM phase transition of the first order.\cite{Klingner2009,Klingner2010}\\
In \yrs\, the very low $T_{N}$ and the itinerant character of the magnetism made it possible to study the unconventional behavior of the thermodynamic and transport properties close to a quantum critical point (QCP).~\cite{Gegenwart2002,Gegenwart2007}
The anomalous behavior is caused by quantum critical fluctuations which may, in principle, be observed directly in the magnetic response $S(\textbf{Q},\omega)$, measured by inelastic neutron scattering.~\cite{Schroeder2000}
These experiments require a certain knowledge of the magnetically ordered structure below $T_{N}$, e.g., the ordering wave vector $\textbf{Q}$. In \yrs\, the structure of the ordered phase is still unknown. 
This is mainly due to experimental problems: limited size of the single crystals ($V\approx 1$\,mm$^{3}$), low $T_{N}$ and also an extremely small ordered moment ($2 \cdot 10^{-3}\mu_{B}$).~\cite{Ishida2003} To overcome these difficulties,
neutron scattering experiments should be carried out under pressure, since $T_{N}$ and the corresponding
ordered moment increase with increasing pressure in Yb-based compounds.~\cite{Knebel2006,Mederle2001} Such experiments are, however, rather difficult.
Isoelectronic substitution of Rh by Co leads to a similar effect as pressure.~\cite{Westerkamp2008,Friedemann2009}
The crystal growing process has been recently optimised in order to produce single crystals of \yrcs, which all crystallize in the tetragonal ThCr$_{2}$Si$_{2}$-type structure: Several high-quality
single crystals with a Co content $x$ varying between 0.03 and 1 have been synthesized.~\cite{Klingner2009,Klingner2010} As expected, increasing $x$ stabilizes magnetic order, enhancing $T_{N}$ and the value of the ordered moment. In this contribution
we present an investigation of the magnetic properties of \ycs\, by means of dc-magnetization measurements. We have derived the low-temperature $H - T$ magnetic phase diagram by applying the magnetic field along three 
crystallographic directions: $[100]$, $[110]$ and $[001]$. Knowledge of the magnetic structure in \ycs\, may help to identify the ordered state in \yrs.\\
The first sign of magnetic order in \ycs\,was observed by $^{170}$Yb M\"ossbauer spectroscopy experiments on polycrystalline materials.~\cite{Hodges1987} AFM order was found below 1.7\,K with the easy magnetization in the basal plane
and a saturated moment of $1.4~\mu_{B}$/Yb, which agrees very well with our preliminary magnetization measurements along the [100] direction.~\cite{Pedrero2010a}
These results could be explained in terms of a Yb$^{3+}$ valence state experiencing a tetragonal crystalline electrical field (CEF), resulting in a Kramers-doublet $\Gamma_{7}$ as the ground state. This agrees well with high-temperature ($T > 100$\,K) susceptibility measurements,
which show a Curie-Weiss behavior with an effective moment of $4.7~\mu_{B}$/Yb and a Weiss temperature $\theta_{W}$ of -4\,K and -160\,K for the magnetic field $H$ parallel and perpendicular to the basal plane, respectively. No magnetic contribution from Co has been
observed.~\cite{Klingner2009,Klingner2010a} A similar fit at temperatures between 2 and 4\,K leads to a reduced effective moment of $3.6~\mu_{B}$/Yb. Neutron scattering experiments revealed that the excited Kramers doublets
are 4, 12.5 and 30.5\,meV away from the ground state doublet.~\cite{Goremychkin2000} An exact analysis of the CEF level scheme has been proposed in Ref.~\onlinecite{Klingner2010a},
which leads to an estimation of the saturation magnetization $M_{s}$ of $1.4\mu_{B}$ and $0.77\mu_{B}$ for $H$ perpendicular and parallel to the crystallographic axis $c$, respectively. A second, clearly first order, AFM phase transition has been observed at
a temperature $T_{L} = 0.9$\,K by means of magnetization, resistivity and specific-heat measurements.~\cite{Pedrero2010a,Mufti2010,Klingner2010a} Both transition temperatures are suppressed by a field larger than 2\,T,
and the $H - T$ magnetic phase diagrams become very complex, showing a strong basal anisotropy.~\cite{Mufti2010} Finally, powder neutron diffraction studies, performed in the intermediate and low-$T$ AFM states, have detected magnetic peaks and suggested
that $\textbf{Q}$ changes its periodicity between the two phases.~\cite{Kaneko2010} In addition, preliminary neutron scattering experiments on single crystals seem to have identified $\textbf{Q}$ and to indicate that $\textbf{Q}$ is incommensurate
below $T_{N}$ and becomes commensurate below $T_{L}$.~\cite{Mufti2010b}\\
Our data can be used to test the CEF calculations by providing the values of the saturation magnetization and clarify the evolution of the magnetic structure in an applied magnetic field.
\section{Experimental}
The single crystals investigated here have been grown with In-flux technique as fully described in Ref.~\onlinecite{Klingner2010a}. They have a residual resistance ratio of about 2 at 0.35\,K and a mass between 8 and 35\,mg. Their shape is square,
with sides parallel to the crystallographic direction [100]. The dc-magnetization ($M$) was measured with a high-resolution Faraday magnetometer, in magnetic fields as high as 3\,T and temperatures down to 0.05\,K.~\cite{Sakakibara1994}
Owing to the magnetic anisotropy of \ycs\,(a factor of about 4),~\cite{Klingner2009,Klingner2010a} the sample platform was modified to reduce the torque contribution of the raw signal, which is proportional to the magnetization perpendicular to the
applied field.
\begin{figure}[b]
\centering
\includegraphics[width=0.45\textwidth]{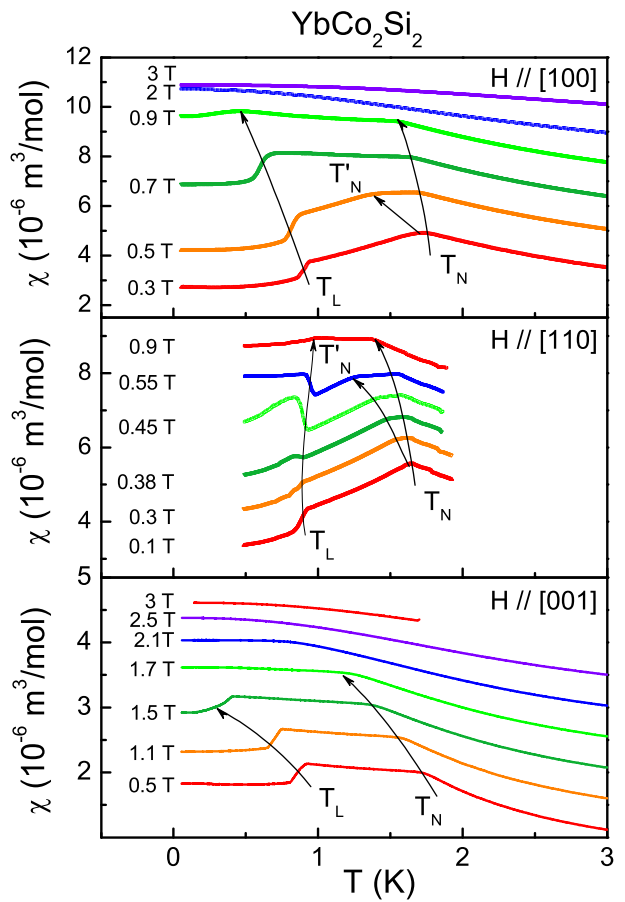}
\caption{Magnetic susceptibility $\chi = M/H$, plotted as a function of $T$, in external magnetic fields applied along three crystallographic directions. The curves
at different fields have been offset for clarity. $T_{N}$ denotes the upper AFM transition temperature, $T_{L}$ the lower one. $T_{N}'$ defines the temperature down to which the susceptibility stays constant below $T_{N}$, thereafter decreasing.
The arrows indicate the evolution of the transition temperatures with increasing magnetic field.}
\label{fig1}
\end{figure}
The platform consists of Stycast epoxy resin. Its small, diamagnetic response to a magnetic field was subtracted. The ac-susceptibility measurements were performed in a Quantum Design Physical Properties Measurement System (PPMS)
in temperatures down to 2 K and magnetic fields up to 3\,T. Some measurements have been carried out down to 0.5\,K in a $^{3}$He option (iQuantum Corporation) for a 7T-SQUID (Quantum Design).

\section{Results}
\subsection{Magnetization vs. temperature}
Figure~\ref{fig1} shows the temperature dependence of the uniform susceptibility $\chi = M/H$ in several fields along the three crystallographic directions. Two features can be identified in all three field directions: a sharp kink at $T_{N}=1.75$\,K and
a distinct drop at $T_{L}=0.9$\,K. At $T_{N}$ AFM order sets in and assumes a different antiferromagnetically ordered structure below $T_{L}$, as suggested in Ref.~\onlinecite{Kaneko2010}. The shape of the magnetization curves is similar
to that observed in \yrs\, (see Ref.~\onlinecite{Schuberth2009}), pointing to an AFM nature of the phase transition at $T_{L}$ in \yrs.
%
%\begin{figure}[b]
%\centering
%\includegraphics[width=0.45\textwidth]{fig2}
%\caption{ZFC (open symbols) and FC (solid symbols) susceptibility for three selected fields along the $[100]$ direction. Labels are described in the caption of Fig.~\ref{fig1}.}
%\label{fig2}
%\end{figure}
%
While the phase transition at $T_{N}$ is second order, the sudden drop at $T_{L}$ and the latent heat observed in the heat capacity in zero field (see Ref.~\onlinecite{Klingner2010a}) point to the first order nature of this phase transition. The entropy above both
transitions confirms a Kramers doublet as the ground state and the local character of the Yb $4f$ quasi-hole. The Kondo temperature has been estimated to be lower than 1\,K.~\cite{Klingner2009,Klingner2010} Both $T_{N}$ and $T_{L}$ shift to
lower temperatures with increasing $H$ along the $[100]$ and $[001]$ directions, whereas for $H // [110]$ $T_{L}$ remains almost constant in $T$. For $H$ parallel to [100] and [110] and at fields higher than 0.1\,T, the sharp cusp at $T_{N}$ changes
into a plateau, where $\chi(T)$ remains almost constant down to a temperature $T_{N}'$ (cf. curve at 0.5\,T for $H // [100]$ and curve at 0.55\,T for $H // [110]$ of Fig.~\ref{fig1}). The lower transition becomes broader in $T$ as the external field is enhanced, and
it disappears for fields $\mu_{0}H \ge 1$\,T along the two directions $[100]$ and $[110]$, but a field larger than 1.5\,T is necessary to suppress $T_{L}$ along $[001]$. A field $H \ge 1.9$\,T is necessary to suppress $T_{N}$ to zero, where
$\chi(T)$ becomes nearly constant, in all three directions.\\
The strong basal anisotropy observed in magnetoresistance measurements (see Ref.~\onlinecite{Mufti2010}) is confirmed by the different behavior of the isofield curves taken along $[100]$ and $[110]$. The curves look very similar for $\mu_{0}H \leq 0.3$\,T but at higher
fields the susceptibility along $[100]$ keeps decreasing below $T_{L}$, while along $[110]$ it increases steeply.
\begin{figure}[t]
\centering
\includegraphics[width=0.45\textwidth]{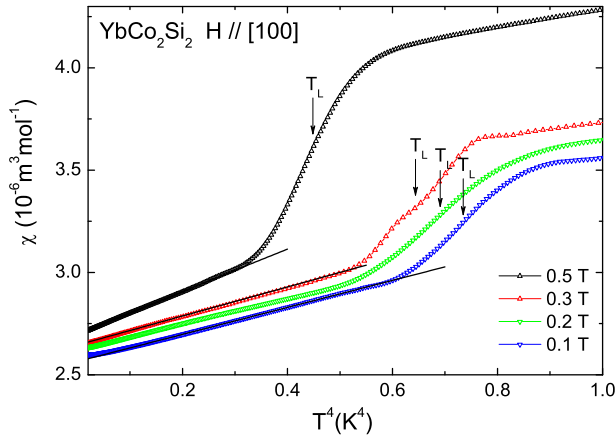}
\caption{Uniform susceptibility plotted as a function of $T^{4}$ to emphasize the behavior below the lower AFM transition temperature $T_{L}$ (indicated by arrows).}
\label{fig3}
\end{figure}
The scenario proposed in Ref.~\onlinecite{Mufti2010} is that the phase transition at $T_{L}$ is the one where the propagation vector $\textbf{Q}$ changes from incommensurate to commensurate. On the other hand, the kink at $T_{N}'$ and the features observed for $H // [110]$
at $\mu_{0}H \approx 0.4$\,T in the isothermal measurements would indicate a possible reorientation of the moments without any change in $\textbf{Q}$. This could explain why $\chi$ changes behavior above 0.4\,T.\\
It is worth noting that, below $T_{L}$ and in low fields parallel to $[100]$, the susceptibility follows a $T^{4}$ dependence, as shown in Fig.~\ref{fig3}. This temperature dependence is due to spin-wave excitations and depends strictly on their
dispersion, which is as yet unknown in this system.
\begin{figure}[t]
\centering
\includegraphics[width=0.45\textwidth]{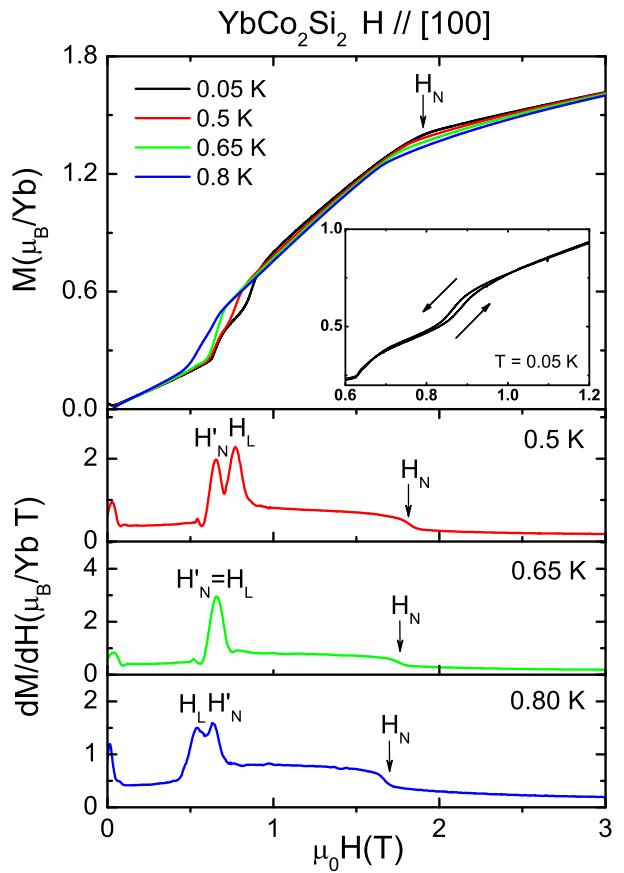}
\caption{Upper panel: $M$ vs. $H$ at four selected temperatures with $H // [100]$. $H_{N}$ denotes the critical field associated with the transition at $T_{N}$. In the inset we zoom on the metamagnetic-like transitions to emphasize the hysteresis loop
at 0.05\,K; the arrows indicate the field sweep direction. In the lower panels, the derivative of the isothermal magnetization is plotted at 0.5, 0.65 and 0.8\,K. The two metamagnetic-like transitions visible at 0.5\,K join each other to become one at
0.65\,K and split again at 0.8\,K. This feature is clearer in $dM/dH$ (bottom panels). We have associated the fields $H_{N}'$ and $H_{L}$ with the transitions at $T_{N}'$ and $T_{L}$, respectively.}
\label{fig4}
\end{figure}
However, calculations based on the low-$T$ propagation vector~\cite{Mufti2010b} can be performed to check this power law, as well as those obtained in specific-heat and resistivity measurements.~\cite{Klingner2010a,Mufti2010}\\
\subsection{Magnetization vs. field}
\begin{figure}[t]
\centering
\includegraphics[width=0.45\textwidth]{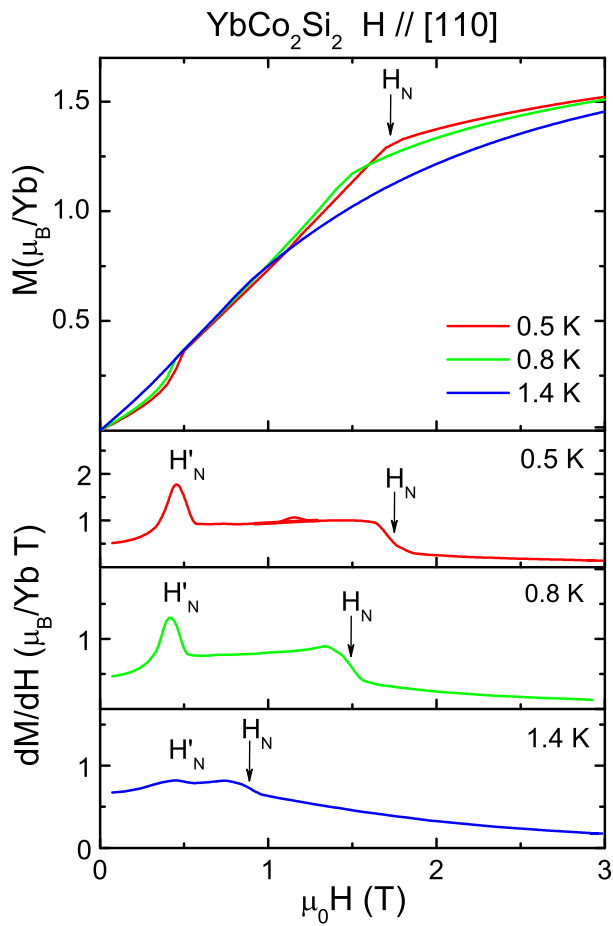}
\caption{Upper panel: $M$ vs. $H$ at three selected temperatures with $H // [110]$. Lower panels: $dM/dH$ vs. $H$ at the same three temperatures. $H_{N}$ and $H_{N}'$ denote the critical fields associated with the transitions at $T_{N}$ and $T_{N}'$,
respectively.}
\label{fig5}
\end{figure}

\begin{figure}[t]
\centering
	\includegraphics[width=0.45\textwidth]{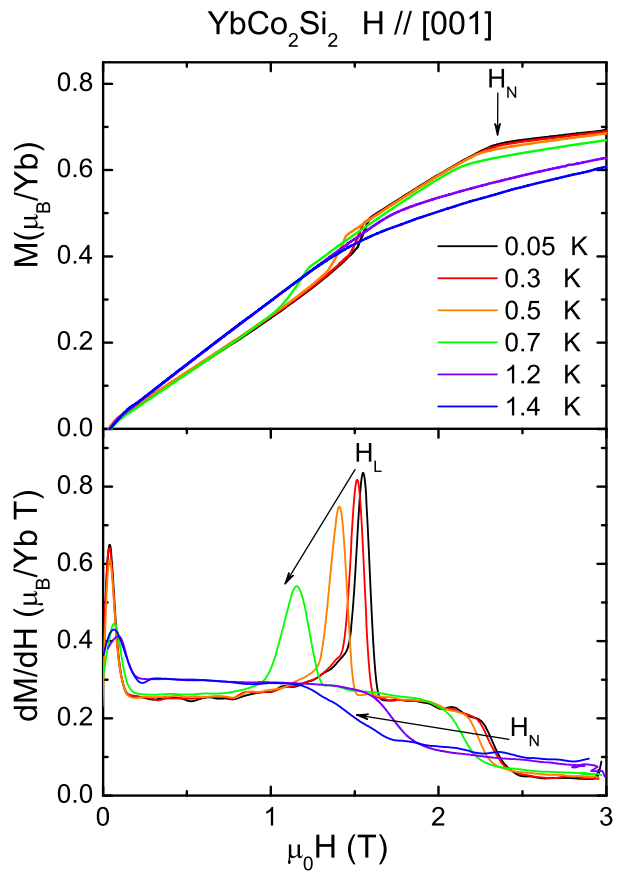}
\caption{Upper panel: $M$ vs. $H$ at six selected temperatures with $H // [001]$. Lower panels: $dM/dH$ vs. $H$ at the same six temperatures. $H_{N}$ and $H_{L}$ denote the critical fields associated with the transitions at $T_{N}$ and $T_{L}$, respectively.
The arrows indicate the evolution of the critical fields with increasing temperature.}
\label{fig6}
\end{figure}

To investigate the phase transition lines in more detail, we measured the field dependence of $M$ at different temperatures. The results are shown in figures~\ref{fig4} to~\ref{fig6} with the respective derivatives $dM/dH$. We start by describing
the data collected with $H // [100]$ (Fig.~\ref{fig4}). The isothermal curves at 0.05\,K show two metamagnetic-like steps at about 0.65\,T and 0.85\,T, followed by a kink at $\mu_{0}H_{N} = 1.9$\,T, which is associated with the transition at $T_{N}$
(upper panel). Measuring $M$, while sweeping the field up and down, a tiny hysteresis is observed across the first two steps that is more pronouced across the one at 0.85\,T (cf. inset of the same panel). At $H_{N}$ the transition is continuous.
At a slightly higher temperature $T \approx 0.2$\,K, the hysteresis vanishes. By further increasing $T$, the two steps merge into one at 0.65\,K and afterwards split again at 0.8\,K without showing any hysteresis.
This is well illustrated in the bottom panels of Fig.~\ref{fig4}, where $dM/dH$ is plotted as a function of $H$. We have assigned the critical fields of the metamagnetic-like transitions to the fields corresponding to the maxima of $dM/dH$ and for the
transition at $H_{N}$ we considered the inflection points in $dM/dH$. Moreover, from the evolution of these anomalies in field we have associated the fields of the first and second step with the signatures seen at $T_{N}'$ and $T_{L}$, namely $H_{N}'$
and $H_{L}$.\\
These signatures indicate that the nature of the phase line $T_{N}'(H)$ may be of the second order while the one of $T_{L}(H)$ is of the first order. This do not contradict the interpretation given in Ref.~\onlinecite{Mufti2010} that at 0.65\,T
the field modifies the orientation of the moments (either through a spin-flop transition or depopulating unfavoured AFM domains, that causes the strong change of slope in $M$ vs. $H$) without modifying the propagation vector $\textbf{Q}$, whereas at 0.85\,T it is the change of $\textbf{Q}$,
that causes the tiny hysteresis effect. This scenario seems to be confirmed by neutron scattering experiments on single crystals performed in magnetic field.~\cite{Mufti2010b}\\
The value of the magnetization just above $H_{N}$ is 1.4\,$\mu_{B}/$Yb, which is in agreement with the saturated moment calculated by Hodges~\cite{Hodges1987} and recently by Klingner~\et,~\cite{Klingner2010a} confirming the local character
of the Yb $4f$ quasi-holes. Above $H_{N}$, $M$ increases further because of van Vleck contributions due to mixing of higher CEF levels.~\cite{Kutuzov08}\\
The results of the $M$ vs. $H$ measurements with $H$ along the $[110]$ direction are shown in Fig.~\ref{fig5}.
At all temperatures lower than $T_{N}$, a metamagnetic-like step is observed at almost the same field of 0.45\,T with no detectable hysteresis. However, the temperatures at which these data have been taken are higher than 0.05\,K (cf. inset of
Fig.~\ref{fig4}). The expected saturation magnetization of 1.4\,$\mu_{B}/$Yb is achieved at $\mu_{0}H_{N} \approx 1.75$\,T for a temperature of 0.5\,K, below which we did not carry out further measurements. At 1.4\,K the signature of such an anomaly is very weak
and can only be observed in the derivative (cf. bottom panel of Fig.~\ref{fig5}). Since only a metamagnetic-like transition is seen along the $[110]$ direction, one may ask whether this transition represents the reorientation of the moments or a change
in $\textbf{Q}$. Since this feature seems to be present also above $T_{L}$, we tend to associate it with $H_{N}'$.\\
Finally, Fig.~\ref{fig6} shows representative curves of the magnetization as a function of the field (upper panel) and its derivative (lower panel) for $H // [001]$. In these measurements also, only a single step could be discerned before the magnetization
reaches its saturation value of 0.68\,$\mu_{B}/$Yb with a kink at 2.35\,T for $T=0.05$\,K. The saturated moment matches very well with that calculated in Ref.\onlinecite{Klingner2010a}.
\begin{figure}[b]
\centering
\includegraphics[width=0.45\textwidth]{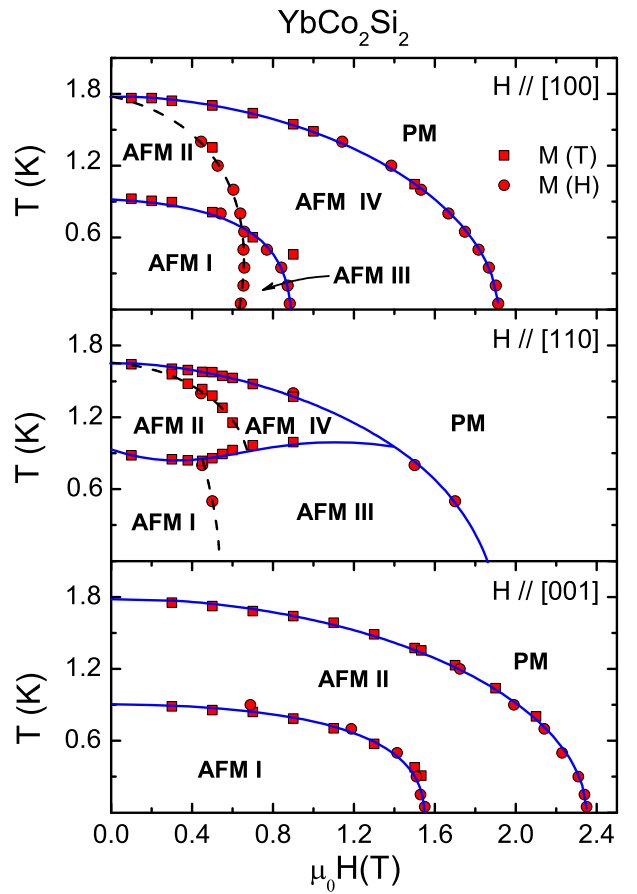}
\caption{Magnetic phase diagram of \ycs\, with $H // [100]$, $[110]$ and $[001]$. The square and circle points have been extracted from isofield and isothermal measurements of $M$, respectively. PM indicates the paramagnetic region while AFM the
antiferromagnetic one. The four different AFM phases are labeled from I to IV. The full lines represent phase transition lines at which the propagation vector $\textbf{Q}$ changes, whereas the dashed lines mark the reorientation of the moments.
The border between the PM and the AFM regions has been fitted with the empirical curve: $[H_{N}(T)/H_{N}(0)]^{n}+[T/T_{N}(0)]^{n} = 1$, where $T_{N}(0)=1.77$\,K, $\mu_{0}H_{N}(0)=1.9$, 1.88 and 2.35\,T and $n$ = 1.9, 1.78 and 1.94 for $H // [100]$, $[110]$
and $[001]$, respectively.}
\label{fig7}
\end{figure}
In this case we have associated the metamagnetic-like step in $M$ vs. $H$ with the phase transition at $T_{L}$, although here no hysteresis could be resolved. This association derives directly from the shape of the phase lines plotted in Fig.~\ref{fig7}.
Both the isothermal and isofield measurements show features at the same phase transition lines where the propagation vector $\textbf{Q}$ is supposed to change. In the lower panel of Fig.~\ref{fig6} we have plotted $dM/dH$ vs. $H$ to emphasize that
the signatures at $H_{L}$ and $H_{N}$ broaden while the peak intensity is reduced, as expected from the evolution of the phase transition lines.
\section{$H~-~T$ phase diagram}
The deduced magnetic $H~-~T$ phase diagram with $H$ applied along the three crystallographic directions $[100]$, $[110]$ and $[001]$ is shown in Fig.~\ref{fig7}. The squares and circles indicate anomalies observed in $M$ vs. $T$ and $M$ vs. $H$, respectively.
The outer second order phase-transition boundary line, which separates the AFM from the paramagnetic (PM) phase, can be followed from 1.75\,K in zero field up to the critical fields $\mu_{0}H_{N}(0)=1.9$, 1.88 and 2.35\,T along the three directions.
The data along these lines can well be described by an empirical curve $[H_{N}(T)/H_{N}(0)]^{n}+[T/T_{N}(0)]^{n} = 1$ with $n$ close to 2. The exponents are displayed in the caption of Fig.~\ref{fig7}. For $H // [110]$
$T_{N}$ at zero field is slightly smaller than that measured along $[100]$ or $[001]$. This is due to the fact that the isofield measurements were performed in the SQUID on a second sample of the same batch.~\cite{Klingner2009}\\ 
Inside the magnetic phase, four AFM regions can be identified when the field is applied along the basal plane (upper panels), while for $H // [001]$ only two regions have been observed (lower panel). We start our description from the lower
panel of Fig.~\ref{fig7}. Since the two AFM phase transitions at $T_{N}$ and $T_{L}$ in zero field have been established to be of the second and first order and involve a change of the propagation vector $\textbf{Q}$,~\cite{Klingner2010a,Kaneko2010,Mufti2010b}
it is straightforward to draw continuous lines on the points and consider the two regions AFM I and II as regions with different $\textbf{Q}$. On the other hand, for $H // [100]$ the $T_{N}(H)$ line seems to split in field separating the regions AFM II
and IV and AFM I and III by a second-order-like phase transition indicated by a kink at $T_{N}'$ or a metamagnetic-like step at $H_{N}'$ without hysteresis (see Figs.~\ref{fig1} and~\ref{fig4}). In addition, the boundaries between the phases
I and II and between the phases III and IV appear to be first order lines (see Fig.~\ref{fig4}). For this reason we have drawn a continuous line, which we think separates the regions with different $\textbf{Q}$. A similar interpretation can be considered
for $H // [110]$, but in this case the line separating the two magnetic structures is almost constant in temperaure. Recent neutron scattering experiments
performed in magnetic field seem to support such a scenario.~\cite{Mufti2010b} Taking into account all our findings
and the results of Refs.~\onlinecite{Mufti2010,Kaneko2010,Mufti2010b} we might interpret our data as follows: Region II is characterized by AFM order with an incommensurate arrangement of the moments, which then assumes a commensurate structure
in region I through a first order phase transition. By applying a magnetic field along $[001]$, the arrangement of the moments undergoes a metamagnetic-like transition
at $T_{L}(H)$ changing the propagation vector and becoming fully polarized above 2.35\,T.
The same feature is expected across the continuous line inside the magnetic ordered phase for $H // [100]$ (upper panel of Fig.~\ref{fig7}),
e.g. betweeen the AFM III and AFM IV. For $H // [100]$ and $[110]$ the magnetic field modifies the orientation of the moments at $T_{N}'(H)$ (dashed lines in Fig.~\ref{fig7}),
either through a spin-flop transition or depopulating unfavoured AFM domains, without modifying the propagation vector $\textbf{Q}$. However, in Ref.~\onlinecite{Mufti2010b} a
significant difference between the behavior of the magnetoresistance across the dashed lines for temperatures lower and higher than $T_{L}(H)$ was observed: For $T > T_{L}(H)$
the magnetoresistance shows a kink at $T_{N}'(H)$ suggesting a continuous transition, e.g. a depopulation of unfavoured AFM domains, while for $T < T_{L}(H)$ the magnetoresistance
shows a distinct drop inferring a spin-flop transition, possibly into a structure with a ferromagnetic component. 
The fact that only for $H // [001]$ no reorientation of the moments is observed, suggests that the moments may lie in directions close to the basal plane, as previously suggested by Hodges.~\cite{Hodges1987}
Our interpretation assumes that the AFM structure allows multiple domains and that the dashed lines in Fig.~\ref{fig7} may indicate a simple domains depopulation effect.
Another and more suggestive possibility is that of a double-$\textbf{Q}$ structure as has been found, e.g., in GdNi$_{2}$B$_{2}$C.~\cite{Jensen2008}
The same principle holds for such a kind of structure in which the field somehow favors one or the other propagation vector.
\section{Ac-susceptibility}
\begin{figure}[t]
\centering
\includegraphics[width=0.45\textwidth]{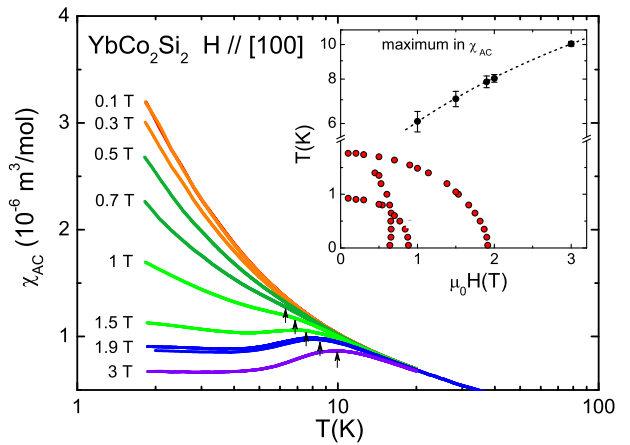}
\caption{Temperature dependence of the ac-susceptibility $\chi'(T)$ measured at different external magnetic fields. The arrows indicate the maximum observed only at high fields. Inset: Magnetic phase diagram with $H // [100]$ in which we have included
the points extracted from the maxima of $\chi'(T)$. The temperature scale was cut to optimize the drawing space.}
\label{fig8}
\end{figure}

Finally, we would like to briefly compare ac-susceptibility $\chi'(T)$ measurements on \ycs\,with those performed on \yrs.~\cite{Westerkamp2008} In the latter compound a maximum was observed in the temperature dependence of $\chi'(T)$ in magnetic field
and it was associated with an energy scale $T^{*}(H)$ interpreted as the energy where the Kondo effect breaks down due to the presence of a field-induced QCP at $H_{N}$. The $T^{*}(H)$ line vanishes for $T \rightarrow 0$ at the QCP.~\cite{Gegenwart2007}
In \ycs~we observe a similar effect, i.e., maxima in $\chi'(T)$ which are indicated by arrows in Fig.~\ref{fig8}. Plotting the maxima on the phase diagram (inset of the same figure) we can deduce from the evolution
of the points that the similar energy scale $T^{*}(H)$ for \ycs\,is not approaching the critical field $H_{N}$. In \ycs\,the maxima undoubtedly represent the thermal excitations of the upper level of the Zeeman split $\Gamma_{7}$ ground state doublet.
Since \ycs\,is a system where the Yb quasi-holes are almost localized while in \yrs\,they are almost delocalized, a straightforward comparison between the two systems cannot be made. However, it would be interesting to study the evolution of this energy
scale while varying the Co content, as has already been done in Refs.~\onlinecite{Westerkamp2008} and~\onlinecite{Friedemann2009}.
\section{Conclusions}
We have explored the magnetic phase diagram of \ycs\,by means of isothermal and isofield magnetization measurements with the magnetic field oriented along the crystallographic directions $[100]$, $[110]$ and $[001]$. In a small
field $\mu_{0}H = 0.1$\,T two AFM phase transitions were detected at $T_{N} = 1.75$\,K and $T_{L} = 0.9$\,K, in the form of a sharp cusp and a sudden drop in $\chi = M/H$. These signatures confirm that the phase transitions are
second order at $T_{N}$ and first order at $T_{L}$. The shape of the magnetization curve is similar to that observed in \yrs\, pointing to the AFM nature of the phase transition at $T_{L}$ in \yrs. The AFM order is completely suppressed by fields
close to 2\,T where the magnetization reaches its saturation values $M_{s}(H // [100]) \approx 1.4 \mu_{B}$ and $M_{s}(H // [001]) \approx 0.68 \mu_{B}$ which match quite well with those calculated for the $\Gamma_{7}$ ground state proposed in
Ref.~\onlinecite{Klingner2010a} and confirm the trivalent state of the Yb ions in \ycs. Inside the AFM phase, two main regions can be identified along all directions where the propagation vector $\textbf{Q}$ assumes two different values.
For $H$ parallel to $[100]$ and $[110]$ (the magnetic structure is anisotropic in the basal plane), however, these regions are separated by another line, which seems to correspond to the line where the magnetic moments reorient, i.e. either
through a spin-flop transition or by changing the domains population; in the case of an AFM double-$\textbf{Q}$ structure we would observe the same features and neutron scattering experiments are needed to shed light on this point. For $H$ parallel
to $[001]$ only two AFM phases have been observed, inferring that the moments likely lay in directions close to the basal plane.
\section{Acknowledgment}
We are indebed to C. Klausnitzer, U. Lie\ss\, and T. L\"uhmann for technical support and K. Kaneko, N. Mufti O. Stockert and M. Rotter for useful discussions.
\bibliographystyle{apsrev}
\bibliography{pedrero}
\end{document}